\documentclass[12pt]{article}

\usepackage{amsmath,amssymb,bm}
\usepackage{epsfig}
\usepackage{graphicx}
\usepackage{amsfonts}
\usepackage{epstopdf}
\usepackage[usenames,dvipsnames]{color} 
\usepackage{float}

\newcommand{\beq}{\begin{equation}}
\newcommand{\eeq}{\end{equation}}
\newcommand{\beqa}{\begin{eqnarray}}
\newcommand{\eeqa}{\end{eqnarray}}

\begin{document}
\def\ii{\'\i}

\title{The effects of a minimal length on the Kerr metric and the Hawking temperature
} 

\author{L. Maglahoui$^{1,2}$, P. O. Hess$^{3,4}$\\
{\small\it $1$ Physics Department, University of Mentouri, Constantine 1}\\
{\small\it Constantine P. O. Box 325, Ain El Bey Way, 025017 Constantine, Algeria}\\
{\small\it $2$ Mathematical and Subatomic Physics Laboratory (LPMPS)}\\
{\small\it University of Mentouri, Constantine 1m Algeria}\\
{\small\it $^3$ Instituto de Ciencias Nucleares, UNAM, Circuito Exterior, C.U.,} \\
{\small\it A.P. 70-543, 04510 M\'exico, D.F., Mexico} \\
{\small\it $^4$ Frankfurt Institute for Advanced Studies,}\\
{\small\it  J. W. von Goethe University, Hessen, Germany}\\
}

\maketitle 

\abstract{A brief review of the {\it pseudo complex General Relativity} (pcGR) will be
presented, with its consequences, as the accumulation of a dark energy
around a mass and a generalized Mach's principle. 
The main objective in this contribution is to determine the Hawking temperature
and the Entropy for various limits: i) The pc-Schwarzschild case with
no minimal length present, ii) the pc-Kerr metric without a minimal length 
and iii) the general case, the pc-Kerr metric with a minimal length present. 
The physical consequences of a minimal length will be discussed, a possible interpretation of a
gravitational Schwinger effect and the appearance of negative temperature. 
For large masses a minimal length does not show any 
sensible effect, but
only for very small masses, several orders of the Planck mass, where non-trivial effects 
emerge, important for the production of mini-black holes
in the early universe. Our results are more general
than being restricted to pcGR. Any other model which assumes a distribution of dark energy
around a stellar body produces the same effects. In contrast to these models,
pcGR demands the presence of a dark energy term. 
}

\section{Introduction}
\label{intro}

In \cite{pcGR-first} the pseudo-complex algebraic extension of the {\it General Relativity} (GR) was
introduced, called {\it pseudo complex General Relativity} (pcGR). The main consequences
of this theory are the mandatory appearance of a dark energy term on 
the right hand side of the Einstein
equation and the existence of a minimal length. 
The accumulation of dark energy around stellar object implies a 
{\it generalized Mach's principle}, namely that any mass not only curves space-time
but also modifies the vacuum structure around it. 

In \cite{book} the theory was explained in detail with the pc-Schwarzschild,
pc-Kerr metric and applications were presented, as to the description of neutron stars
and the accretion disk structure around a black hole. In \cite{PPNP} the pcGR was
put into relation to several algebraic extensions of the GR, showing that all versions
are contained as special cases of pcGR. 
Accretion disks around a black hole were investigated \cite{book,PPNP,accr} and it was 
shown that deviations from GR only appear very near to the event horizon. A dark 
ring in the accretion, followed by a bright one is predicted, however, an observable resolution
of minimal $5\mu as$ are required, not achieved up to now.
Another consequence of this theory is that any mass
can be stabilized, though very heavy objects still resemble black holes as predicted by the GR.
Nevertheless, some possible emissions of light, due to in-falling matter, at/near the poles is
predicted \cite{hess-light-poles}. Therefore, pcGR predicts that stellar objects larger than 2.3
solar masses will be observed which neither behave as neutron stars nor as black holes, 
the transition to stellar objects resembling black holes will be fluent.

In this contribution, we are interested in determining the Hawking temperature and the entropy
of these stellar objects, as a function of the rotational Kerr parameter $a$ and $b$, the last 
describing the accumulation of dark energy around a stellar object. We also propose that all space
outside a stellar object is subject to a Hawking radiation, similar to the {\it Schwinger effect}
known in {\it Quantum Electrodynamics} (QED) \cite{schwinger-effect}. Another interest is 
to study the effects of a minimal length on this radiation.
The pcGR will show that when an event horizon is present, near to this horizon negative
temperatures will appear, which correspond to negative pressure \cite{T-neg}, thus,
stabilizing the star.

The paper is organized as follows: In Section \ref{pcGR} the pcGR is resumed and its main consequences
listed. In Section \ref{kerr} the application to the Kerr metric is discussed in several steps.
First the limit of a non-rotating stellar object with no minimal length is investigated, after that
the Kerr metric with no minimal length is considered and finally the general case, a rotating
stellar object with a minimal length present. The last case will show dramatic effects near
the event horizon
In Section \ref{con} Conclusions 
are drawn.

\section{Pseudo-complex General Relativity}
\label{pcGR}

In this section, we will resume the theory of pcGR, i.e., its motivation, structure and 
some important consequences. The presentation will unite several results which have been published
separately, but are important for a complete picture.

One possible manner to modify GR is an {\it algebraic extension} \cite{PPNP}. With that notion we
associate the use of alternative variables than $x^\mu$, with $\mu$ = 0, 1, 2, 3. However, not all algebraic 
extensions are consistent, as noted in \cite{kelly}: Most algebraic extensions lead in the
limit of weak gravitational fields to the existence of ghost  and/or tachyon solutions. Only two
sets of coordinates are without problems, 
namely the real coordinates $(x^\mu )$, which are the standard coordinates in GR,
and the pseudo-complex coordinates

\beqa
X^\mu & = & x^\mu + I y^\mu
~~~,
\label{pcGR-1}
\eeqa
with $I^2 = 1$ and $y^\mu$, yet to be determined. Thus, the space in pc-GR is 8-dimensional
and one has to find a procedure to reduce it again to a 4-dimensional space, which will be 
exactly what will be explained further below through the use of a constraint.

Alternatively to (\ref{pcGR-1}) the $X^\mu$ can be expressed in terms of newly defined operators
$\sigma_\pm = \frac{1}{2}\left( 1 \pm I \right)$, i.e.,

\beqa
X^\mu & = & X^\mu_+ \sigma_+ + X^\mu_- \sigma_-
\nonumber \\
\sigma_\pm^2 & = & \sigma_\pm ~,~ \sigma_+ \sigma_- = 0 ~,~ \sigma_+ + \sigma_- ~=~ 1 ~,~
\sigma_+ - \sigma_- ~=~ I
\nonumber \\
X^\mu & = & \frac{1}{2}\left( x^\mu \pm y^\mu \right)
~~~,
\label{pcGR-2}
\eeqa
which provides a division of $X^\mu$ in terms of the {\it zero-divisor components} of $\sigma_\pm$.
The notion of a {\it zero divisor} is a consequence of the property $\sigma_+\sigma_- =0$,
i.e., two numbers $\lambda_+\sigma_+$ and $\lambda_-\sigma_-$ multiplies gives zero,
which shows that the pseudo-complex variables do not form a field
but rather a ring.

The length element squared in pcGR has now the form

\beqa
d\omega^2 & = & g_{\mu\nu}(X) dX^\mu dX^\nu
\nonumber \\
& = & g^+_{\mu\nu}(X_+) dX_+^\mu dX_+^\nu \sigma_+ + g^-_{\mu\nu}(X_-) dX_-^\mu dX_-^\nu \sigma_-
\nonumber \\
& = & \left\{ g^+_{\mu\nu}(X_+) dX_+^\mu dX_+^\nu + g^-_{\mu\nu}(X_-) dX_-^\mu dX_-^\nu \right\}
\nonumber \\
&& + I \left\{ g^+_{\mu\nu}(X_+) dX_+^\mu dX_+^\nu - g^-_{\mu\nu}(X_-) dX_-^\mu dX_-^\nu \right\}
\nonumber \\
& = &
\left\{g^S_{\mu\nu}\left[ dx^\mu dx^\nu +
dy^\mu dy^\nu \right]
+ g^A_{\mu\nu}\left[ dx^\mu dy^\nu + dy^\mu dx^\nu\right]\right\}
\nonumber \\
&& 
+ I \left\{ g^A_{\mu\nu} \left[ dx^\mu dx^\nu + dy^\mu dy^\nu
\right] +
g^S_{\mu\nu} \left[ dx^\mu dy^\nu + dy^\mu dx^\nu \right]
\right\}
~~~,
\label{pcGR-3}
\eeqa
with the symmetric and antisymmetric combinations of the metrics from the two
zero divisior components:

\beqa
g^S_{\mu\nu} & = & \frac{1}{2}\left(
g^+_{\mu\nu} + g^-_{\mu\nu} \right)
~,~ 
g^A_{\mu\nu} ~ = ~ \frac{1}{2}\left(
g^+_{\mu\nu} - g^-_{\mu\nu} \right)
~~~.
\label{pcGR-4}
\eeqa

The length element has to be real (particles only move on a real trajectory) and, thus, 
demanding that the pseudo-imaginary component in (\ref{pcGR-3}) vanishes, represents a 
{\it constraint} which has to be implemented.

The equations of motion, i.e. the modified Einstein equations, are derived from a variational
principle, at first sight to be identical to the one in GR, but in distinction all objects (tensors)
are {\it pseudo-complex} (pc) \cite{book,PPNP}. The action is defined as

\beqa
S & = &\int dX^4 \sqrt{-g}\left({\cal R}+2\alpha \right)
\label{pcGR-5}
\eeqa
with ${\cal R}$ as the pc-Riemann scalar and the term $\alpha$ describes the
an additional 
dark energy. In the first publication, where the pcGR was presented \cite{pcGR-first,book}, a different variational principle was proposed, which is more involved to implement. But already in
the appendix in \cite{book} the alternative proposal, to apply instead a constraint, 
was discussed.

Therefore, one imposes the variational principle $\delta S = 0$ with a constraint, namely

\beqa
\delta S & = & \delta S_+ \sigma_+ + \delta S_- \sigma_- ~=~  0
\nonumber \\
0 & = & 
\left(\sigma_+ - \sigma_-\right) 
\left\{ g^+_{\mu\nu}(X_+) dX_+^\mu dX_+^\nu - g^-_{\mu\nu}(X_-) dX_-^\mu dX_-^\nu \right\} 
\nonumber \\
& = &
I \left\{ g^A_{\mu\nu} \left[ dx^\mu dx^\nu + dy^\mu dy^\nu
\right] +
g^S_{\mu\nu} \left[ dx^\mu dy^\nu + dy^\mu dx^\nu \right]
\right\}
~~~,
\label{pcGR-6}
\eeqa
varying with respect to $g^+_{\mu\nu}$ in the $\sigma_+$ zero-divisor component and
with respect to $g^-_{\mu\nu}$ in the $\sigma_-$ zero-divisor component. The constraint
is implemented via the use of a Lagrange multiplier $\lambda$ \cite{accr}. 

This variation leads to the modified Einstein equations \cite{PPNP}

\beqa
{\cal R}^\pm_{\mu\nu} - \frac{1}{2}g^\pm_{\mu\nu}{\cal R}_\pm
& = &
8\pi T_{\pm~\mu\nu}^\Lambda
~~~,
\label{pcGR-7} 
\eeqa
with the energy momentum tensors in the zero-divisor basis given by

\beqa
8\pi \kappa T_{\pm~\mu\nu}^\Lambda & = & \lambda u_\mu u_\nu + \lambda \left( {\dot y}_\mu {\dot y}_\nu 
\pm u_\mu {\dot y}_\nu \pm u_\nu {\dot y}_\mu\right) + \alpha g_{\mu\nu}^\pm
~~~,
\label{pcGR-8}
\eeqa
where the dot refers to the derivative with respect to the eigentime ($ds=d\tau$).

To introduce a constraint is paid with a price, namely to find a solution to this constraint.

From this, we conclude that using pc-coordinates leads to Einstein equations which
{\it demands the appearance of a non-zero energy-momentum tensor} on the right hand
side of the equations. At this point, we have to introduce a phenomenological ansatz, which is based on
the observation that a curved space creates vacuum fluctuations \cite{birrell,visser},
resulting in an $\sim 1/r^6$ dependence, times a function in $r$.
As shown in \cite{visser} these vacuum fluctuations diverge at the event horizon, which is
probable a consequence of not taking into account back-reaction effects. Nevertheless, it
shows that vacuum fluctuations are strong near the position of the event horizon and the
dependence on the radial distance we take into account by a fall-off of the dark energy as
$\sim \frac{1}{r^6}$. Lower powers are excluded by observation \cite{PPNP}, but the fall-off can
be stronger. In \cite{PPNP} the fall-off of the dark energy is parametrized as $1/r^n$.

In what follows we will show that after some justifiable approximations we are left with
one set of real Einstein equations, with a non-zero energy-momentum tensor on its right hand side,
and the solution of the constraint will lead to a dispersion relation, which in a flat space
is the one known from special relativity. The two equations in (\ref{pcGR-7}) will, under the 
proposed approximations, fuse to one equation.

These approximations are: i) We assume that the metrics in the two zero-divisor 
component are exactly the same, namely

\beqa
g^\pm (X_\pm ) & \approx & g_{\mu\nu}(x)
~~~.
\label{pcGR-9}
\eeqa
This is equivalent to assume the same functional form of the metrics, i.e.,
$g^\pm (X_\pm ) = g(X_\pm )$. ii) A further assumption is that the corrections to the minimal length
in the metric
are negligible, i.e., the dependence is only on the real variable $x^\mu$. This approximation
also implies that

\beqa
g^S_{\mu\nu} (x) & = & g_{\mu\nu}(x) ~{\rm and}~ g^A_{\mu\nu} (x) ~=~ 0
~~~.
\label{pcGR-10}
\eeqa

With this, the constraint reduces to

\beqa
g_{\mu\nu} (x) dx^\mu dy^\nu & = & 0
~~~.
\label{pcGR-11}
\eeqa
This is the {\it generalized dispersion relation} with the solution for $y^\mu$ given by

\beqa
y^\mu & = & l \frac{D x^\mu}{D\tau}
~~~,
\label{pcGR-12}
\eeqa
with $D$ being the covariant derivative \cite{adler} and $\tau$ is the eigentime. 
The factor $l$ is introduced for dimensional reasons and has the dimension of length ($c=1$).

Here, we have an example of a theory which uses a minimal length as a {\it parameter},
i.e., it is not affected by a Lorentz transformation. All symmetries in GR are preserved in
pcGR, which is a huge advantage.
For the case of a flat space 
($g_{\mu\nu}=\eta_{\mu\nu}$, with $\eta_{\mu\nu}$ the Minkowsky metric) the dispersion relation
reduces to the standard one and the solution of $y^\mu$ is just
$y^\mu = l u^\mu$,
with $u^\mu$ as the 4-velocity. It is obtained directly from (\ref{pcGR-12}), noting
that all Christophel symbols vanish in flat space. 
This is the standard assumption
for $y^\mu$ in theories including a minimal length (maximal acceleration)
\cite{caianiello,feoli}.

Using (\ref{pcGR-10}), the Einstein tensors $G^\pm_{\mu\nu}$, which are pure functions of the metric,
are in both zero divisor component equal, yielding $G_{\mu\nu}$.
The energy-momentum tensors are also set to $T^\Lambda_{\mu\nu}$
= $\frac{1}{2}\left( T^\Lambda_{+~\mu\nu}+T^\Lambda_{-~\mu\nu}\right)$ (exploiting the symmetry in
the zero-divisor components), 
where we added the upper symbol $\Lambda$ in order to refer it to dark energy. 

To resume, we are
finally left with the equations

\beqa
G_{\mu\nu} & = & 8\pi \kappa T^\Lambda_{\mu\nu}
\nonumber \\
y^\mu ~=~ \frac{Dx^\mu}{D\tau} & \leftarrow &  g_{\mu\nu} dx^\mu dy^\mu = 0 
\nonumber \\
d\omega^2 & = & g_{\mu\nu} (x) \left[ dx^\mu dx^\nu + dy^\mu dy^\nu \right]
~~~.
\label{pc-GR-14}
\eeqa

The ansatz for $T_{\mu\nu}$ is made using a phenomenological approach, 
based on results from semi-classical
Quantum Mechanics \cite{birrell,visser} and it was explained above. 

A phenomenological parametrization is finally used, describing a falling-off
of the dark energy density as a power in the radial distance. For a particular ansatz, please
consult \cite{PPNP}

The main results of this section can be resumed as follows:

\begin{itemize}

\item The pc-extension of GR (pcGR) demands a non-zero energy-momentum tensor on the right hand side
of the Einstein equations. This term is associated to the appearance of a vacuum with an internal 
structure, given by a dark energy.  The main consequence is as follows:

\item The pc-GR implies an extension of Mach's principle, namely
that {\it Mass deforms not only space-time} but
{\it also changes the vacuum properties around any mass}. We  denominate it as 
{\it the extended Mach's principle}. 
This result may be important, giving hints 
on what a quantized Gravity has to include.

\item The pcGR also has as a consequence the generalized dispersion relation, resulting
in conditions for the four components of $y^\mu$. Thus, it 
describes how a particle moves in the 8-dimensional space, restricting it to a 
4-dimensional embedded space.

\item Another consequence of the constraint is the appearance of a {\it minimal length parameter} $l$.
This parameter is a scalar and is not affected by a Lorentz transformation. It represents a huge
advantage, compared
to other theories with a physical minimal length,
subject to Lorentz contraction. In addition, the appearance
of a minimal length may shed, too, some light on what a
quantized Gravity has to contain.

\end{itemize}

\section{Application to the Schwarzschild and Kerr metric}
\label{kerr}

Defining a dimensionless radial variable

\beqa
y & = & \frac{r}{m}
~~~,
\label{intro-3}
\eeqa
outside the star's mass distribution, 
the non-zero components of the Kerr metric are \cite{PPNP}

\beqa
g_{00} & = & -\frac{ y^2 - 2 y  + a^2 \cos^2 \vartheta + \frac{b}{6y^{2}} }{y^2 + a^2 \cos^2\vartheta}~~~, \nonumber \\
g_{11} & = & \frac{y^2 + a^2 \cos^2 \vartheta}{y^2 - 2 y + a^2 +  \frac{b}{6y^{2}}  }~~~, \nonumber \\
g_{22} & = &  y^2 + a^2 \cos^2 \vartheta~~~,  \nonumber \\
g_{33} & = & (y^2 +a^2 )\sin^2 \vartheta + \frac{a^2 \sin^4\vartheta \left(2 y -  \frac{b}{6y^{2}}  \right)}{y^2 + a^2 \cos^2 \vartheta}~~~,  \nonumber \\
g_{03} & = & 2 \frac{-a \sin^2 \vartheta ~ 2 y 
+ a \frac{b}{6y^{2}}   \sin^2 \vartheta }{y^2 + a^2 \cos^2\vartheta}
~~~,
\label{intro-2a}
\eeqa
where $b$ is the parameter for the size of accumulation of dark
energy around a stellar object. These components change insside the stars' mass distribution,
which will not be considered here. For a possible path, please consult \cite{caspar}.

In order to calculate the position of the event horizon, we determine the zeros
of the $g_{00}$ component, which is equal to solving the equation \cite{symmetries-2020}

\beqa
y^4 - 2 y^3 + a^2 y^2 + b/6 & = & 0
~~~,
\label{eq-1}
\eeqa
with four solutions.

In what follows, several limits will be discussed, 
as the Schwarzschild solution ($a=0$) but
no minimal length, the Kerr solution but still with $l=0$ and finally the most general case,
including the minimal length. There, we will see that the minimal length has important 
consequences for the structure of small black holes.

\subsection{Schwarzschild case, $a=0$ AND $l=0$}
\label{sec1}

The basis of the following calculations is the 
Schwarzschild metric, obtained using $a=0$ in the Kerr metric (\ref{intro-2a}).

For $b=0$ we arrive at the standard expression in General
Relativity (GR). For $b=\frac{81}{8}$, we arrive at the value used in
the pseudo-complex General Relativity (pcGR)
\cite{pcGR-first,PPNP}.

\subsubsection{Hawking temperature}
\label{sub-1.1}

According to \cite{padman}, the surface gravity $\kappa$,
is obtained via

\beqa
\kappa & = & \frac{1}{2} \frac{\partial g_{00}}{\partial r}
~ = ~ \frac{1}{2m}\frac{\partial g_{00}(y)}{\partial y}
~~~,
\label{temp-2}
\eeqa
where it is important to note that $\kappa$ times the mass $m$ is only a function
in $y$.
The derivative is usually applied at the position of the event horizon.
In \cite{padman} essentially the same method is applied.
Here, we will also calculate the $\kappa$ at an arbitrary distances to
the center. 
The reason lies in our assumption that pair creation may be possible as soon
as there is a gravitational field \cite{visser,falke2023}, of course for weak fields the yield would be
negligible. The same assumption is used in Quantum Electrodynamics (QED) and this
phenomenon is called the {\it Schwinger effect} \cite{schwinger-effect}, which we will denote as
the {\it gravitational Schwinger effect}.

\begin{figure}
\centering
\includegraphics[width=0.40\textwidth]{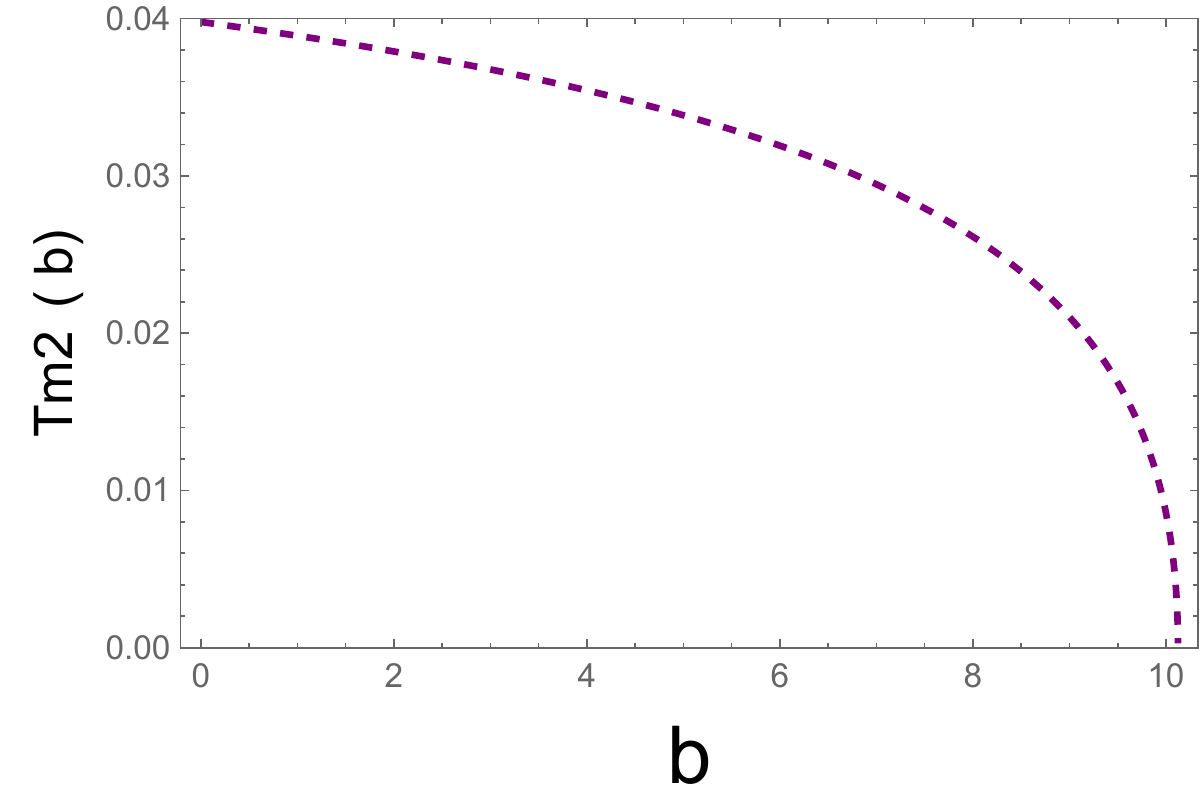} 
\caption{The temperature times $m$ depicted in dependence on $b$ for the
Schwarzschild limit. Note, that the
temperature is approaching zero as $b$ approaches 10.125.
} 
\label{tvsb}
\end{figure}

The Hawking temperature is related to the surface gravity via \cite{birrell,padman}

\beqa
T & = & \frac{\kappa}{2\pi} ~=~ \frac{1}{2\pi m y_h^2} \left( 1- \frac{b}{3y_h^3}\right)
~~~,
\label{temp-4}
\eeqa
where the final result is given on the right hand side. The position $y$ is taken here at the
event horizon. But, note that we also can calculate $T$ at arbitrary distances, where
$y$ is used instead of $y_h$.
In Fig. \ref{tvsb} the temperature at the event horizon is plotted versus b. 
Note, that $T$ times the mass $m$ 
is only a function in $y$, only valid outside the star's mass distribution, i.e.
for $y \geq y_h$, for the inside a different approach has to be applied.
For $y>y_h$ the temperature again is positive.

As a general result, the Hawking temperature starts at the known value 
for $b=0$ (GR) and tends to zero for $b=\frac{81}{8}$. For the last value,
there is no emission from the black hole at the event horizon, implying $T=0$ and
a stable black hole, which does not evaporate. In fact, at this point the
black hole ceases to exist and is a normal dark star, 
{\it no emissions}.   

This strange behavior is linked to the negative heat capacity, which is
obtained deriving $T$ with respect to $m$. Because $T$ is inverse proportional to $m$,
the heat capacity will be proportional to $-1/m^2$. I.e., when $m$ increases,
the entropy increases but the temperature decreases and vice versa.

\subsubsection{Entropy, for the case $a=0$ and $l=0$}
\label{sub-entropy-sch}

The relation between the entropy and energy is given by the
thermodynamical relation  (no electric charge, no angular momentum)

\beqa
dE ~=~ dm & = & TdS
~~~,
\label{entropy-1}
\eeqa
where $m$ is the energy (mass) of the system (note that $c=1$). 

Substituting (\ref{temp-4}) for $T$ leads to the differential equation

\beqa
dS & = &
\left(2 \pi my_h^2\right) dm\left[ 
\left( 1 - \frac{b}{3 y_h^2}\right)  \right]^{-1}
~~~.
\label{entropy-2}
\eeqa

Inspecting (\ref{entropy-2}) the $dS$ is proportional to $m dm$ and it
can be integrated easily, leading to

\beqa
S & = & \frac{m^2}{2}
\left(2 \pi y_h^2\right) \left[ 
\left( 1- \frac{b}{3 y_h^2}\right) \right]^{-1}
~~~.
\label{entropy-3}
\eeqa
Now, the factor $\left( 1- \frac{b}{3 y_h^2}\right)$ 
is in the denominator! This implies that
the entropy goes to infinity toward the event horizon, while $T$ 
goes to zero, which is the result of the simplified approximations applied.
Nevertheless, the product $TdS$ is constant and is always 
given by $dm=dE$, i.e., finite. 

\subsection{The Kerr case for $l=0$}
\label{sec1a}

\begin{figure}
\begin{center}
\rotatebox{0}{\resizebox{400pt}{400pt}{\includegraphics[width=0.23\textwidth]{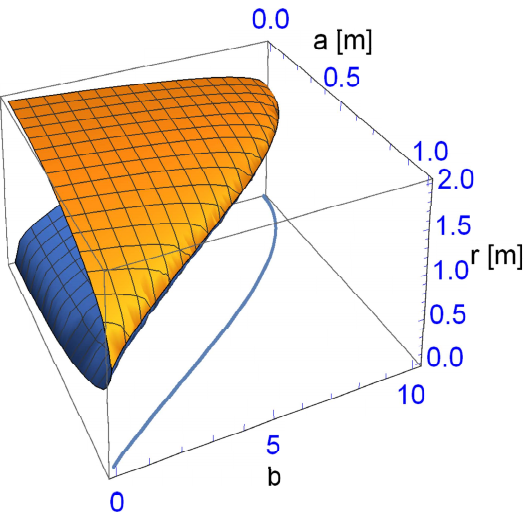}}}
\caption{The position of the event horizon 
$r_h= m y_h$, as a function
on the dark energy parameter $b$ and the rotational paraneter
$a$ \cite{symmetries-2020}. The lower sheet in the figure corresponds to
inner event horizons, also a solution of
(\ref{eq-1}).
\label{figure2}
}
\end{center}
\end{figure}

The metric, in terms of the radial variable $y$, is given in (\ref{intro-2a}). In 
order to obtain the position of the event horizon, the equation (\ref{eq-1}) has to be resolved.
The corresponding figure is depicted in Fig. \ref{figure2}, where the dependence in terms of
$b$ and $a$ is given \cite{symmetries-2020}. The main observations are: For $b=\frac{81}{8}$ there is still an
event horizon at $a=0$. However, as soon as $a>0$ there is no event horizon anymore.
Also, for a given $b$, increasing $a$ leads at some point on to the disappearance of
the event horizon.
The curve at the bottom of Fig. \ref{figure2}
traces the projection of the event horizon, in dependence on $a$ and $b$. 

The Hawking temperature is calculated in the same manner as explained in the last section,
save that now there is an additional dependence on $a$.

\begin{figure}
\begin{center}
\rotatebox{0}{\resizebox{400pt}{400pt}{\includegraphics[width=0.23\textwidth]{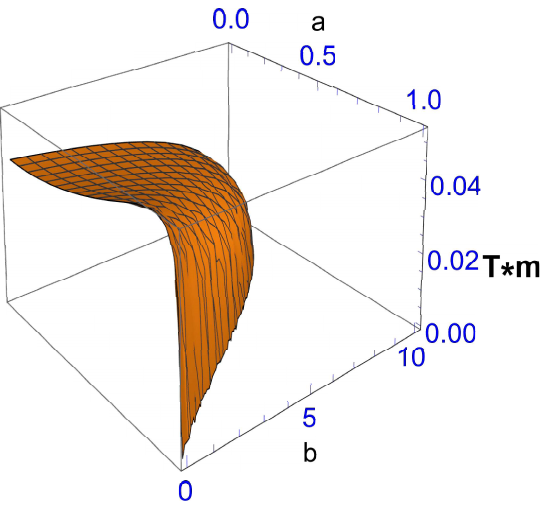}}}
\caption{The dependence of the temperature, multiplies with the mass parameter $m$, on the rotational parameter $a$ of the Kerr metric
and on the dark energy parameter $b$. We used
$\epsilon = \frac{l}{m}=0.01$.
\label{figure3}
}
\end{center}
\end{figure}

For the temperature we obtain

\beqa
T & = & \frac{\kappa}{2\pi}
~ = ~ \frac{1}{2 \pi my_h^2} 
\left( 1- \frac{a^2 {\rm cos}^2(\theta )}{y_h}
- \frac{b}{3 y_h^2}\right) \left( 1 + \frac{a^2
{\rm cos}^2 (\theta )}{y_h} \right)
~~~.
\label{temp-7}
\eeqa
Also, here we set $y=y_h$, but as mentioned, the same formula is valid for any 
position $y$.
In Fig. \ref{figure3} the temperature times the mass $m$ is depicted in dependence on $a$ and $b$.
For $a=0$ the temperature decreases continuously towards zero, as obtained in the former section.

\begin{figure}
\centering
\includegraphics[width=0.40\textwidth]{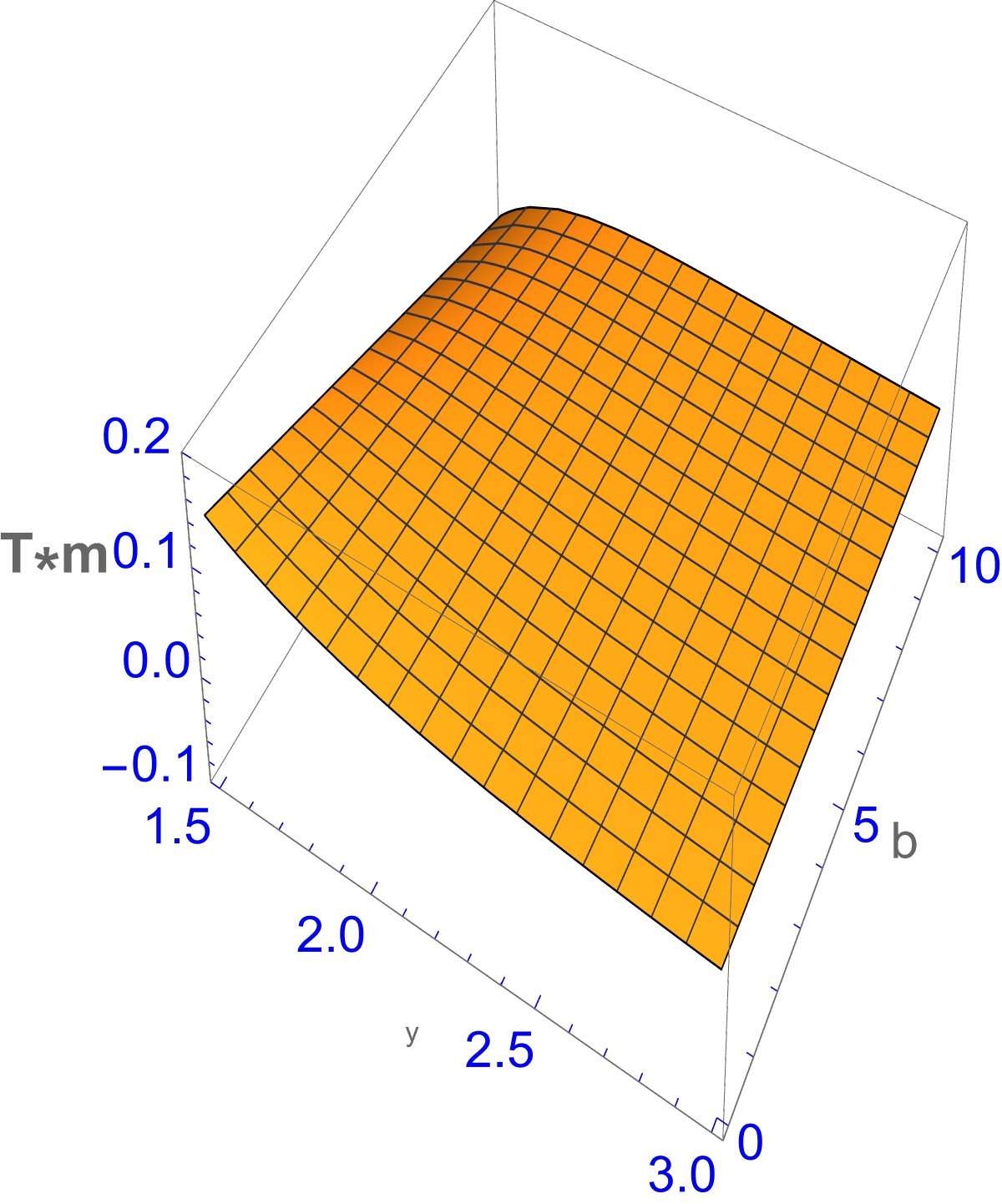}  
\includegraphics[width=0.40\textwidth]{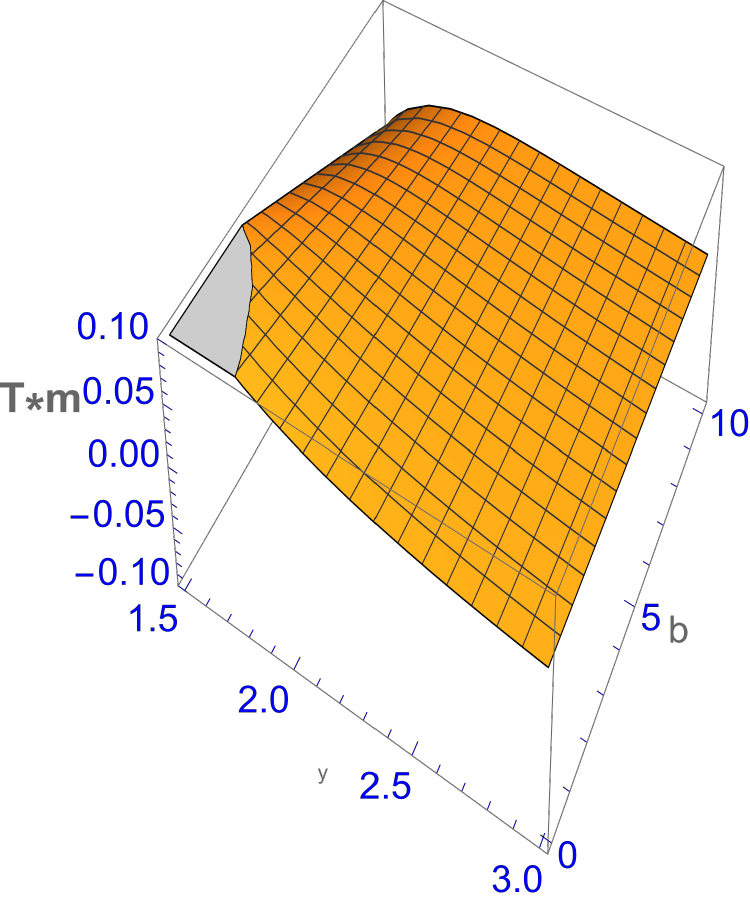}  
\caption{ $T$ times $m$ is plotted versus $y$ and $b$. 
Note, due to the large mass ($\epsilon = \frac{l}{m} = 0$), 
there is no effect of the minimal length.
} 
\label{fig1ab}
\end{figure}

Also interesting is plotting the temperature, times $m$, as a function in $y$ and $b$ for 
two different cases of $a$, namely 0 and 1. The result is given in Fig. \ref{fig1ab}.
In Quantum Electrodynamics particle pairs can be created in strong electromagnetic fields,
which is called the {\it Schwinger effect} \cite{gelis2016}. With the same reason, we can assume the
possibility that pair production can happen in strong gravitational fields,as shown in \cite{visser}. 

\begin{figure}
\centering
\includegraphics[width=0.40\textwidth]{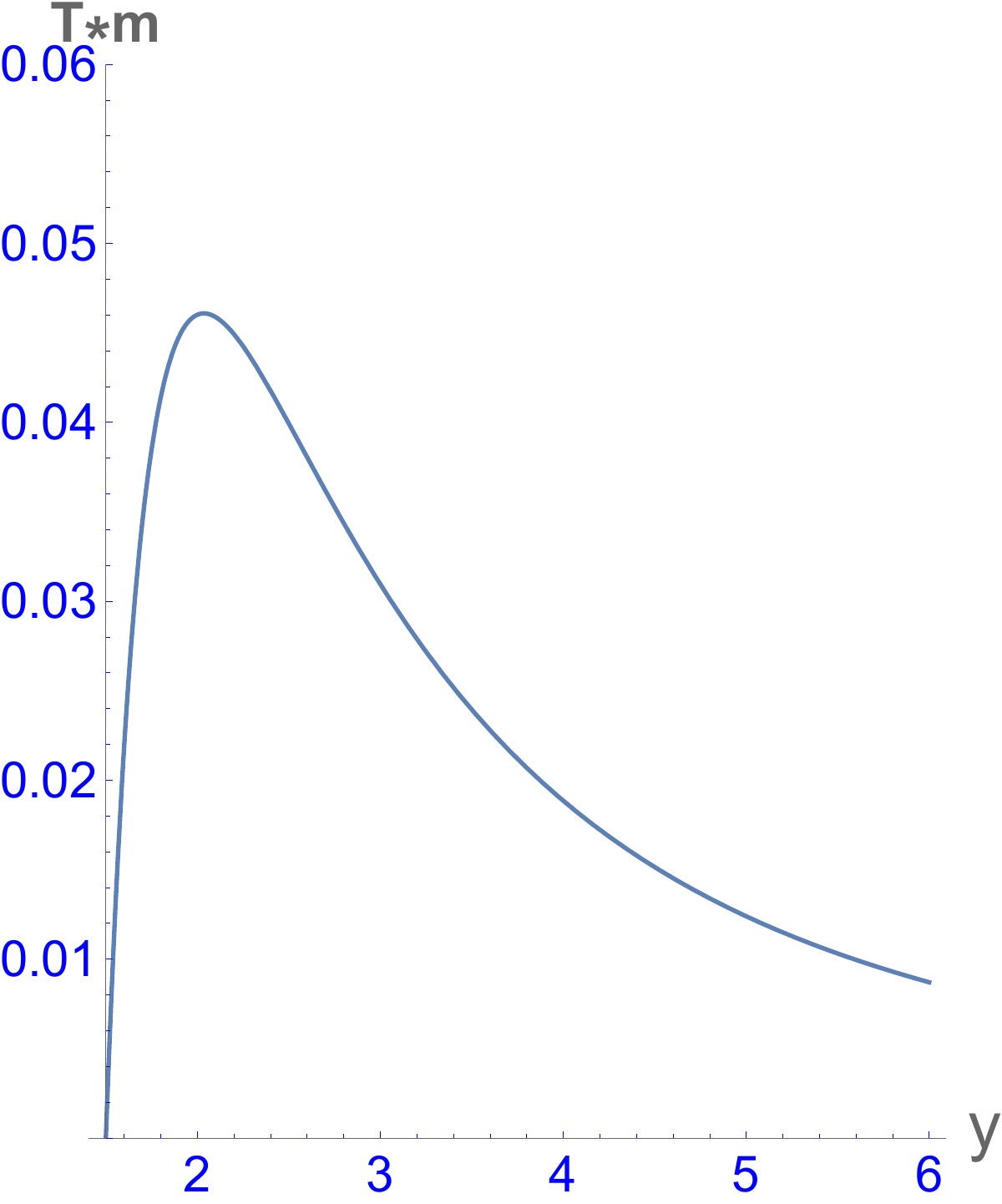}  
\includegraphics[width=0.40\textwidth]{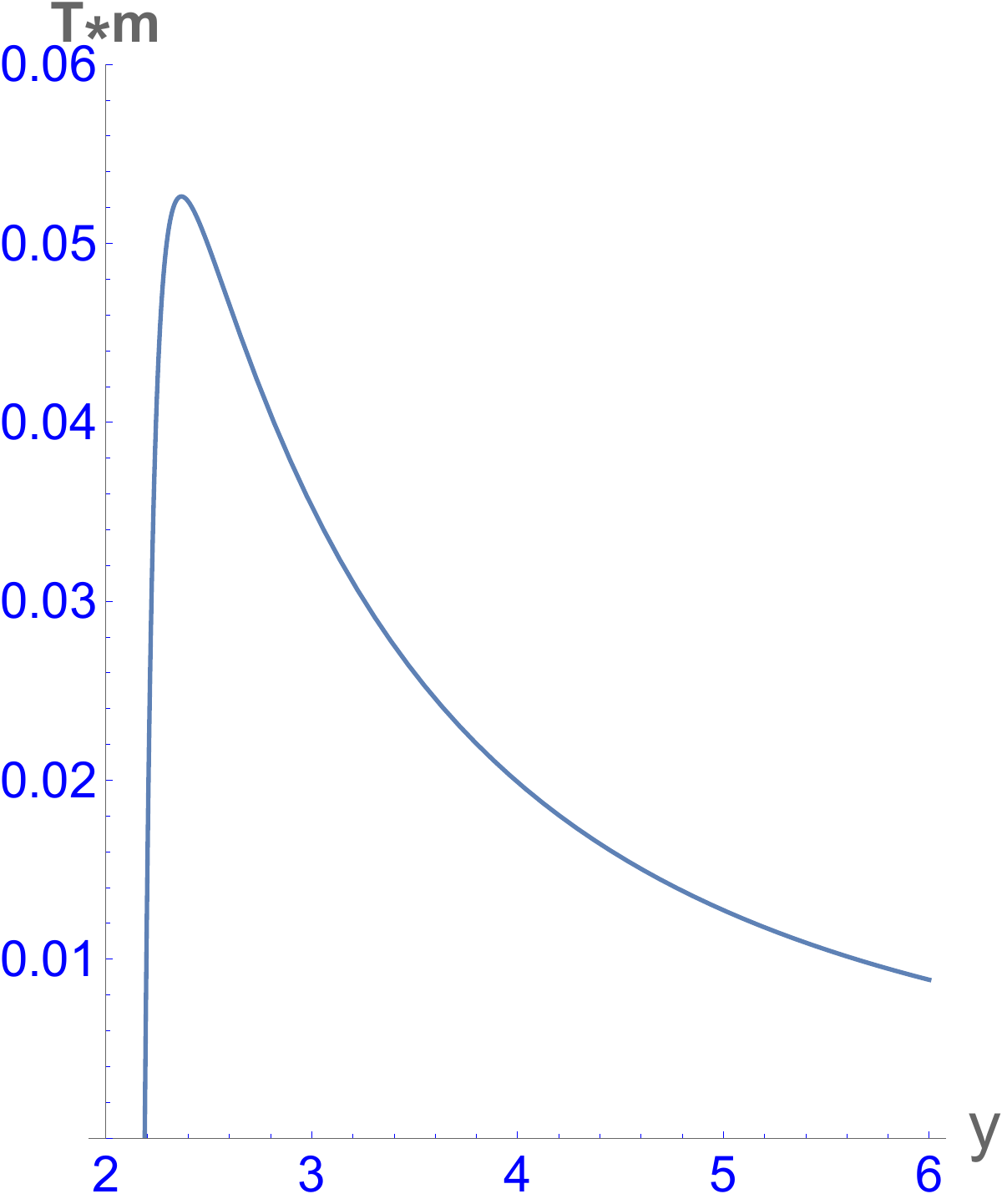}  
\caption{ $T$ times $m$ is plotted versus $y$, for $b=\frac{81}{8}$.
The left panel shows the result without the factor $\sigma^2$ and the
right panel with this factor. The value $\epsilon = 0.01$ was used.
} 
\label{figcut}
\end{figure}

In Fig. \ref{figcut} the function $T m$ is plotted versus $y=\frac{r}{m}$. For 
a non-zero value of $b$, vacuum fluctuations are simulated. The effect of them
is a peak in the function $Tm$, as also obtained in \cite{falke2023}, where the
vacuum fluctuations were calculated using first principles. 
The fact that by our phenomenological ansatz we obtain a similar structure shows the
usefulness of phenomenological approaches in getting useful insights.
The left panel in Fig. \ref{figcut}
is for setting the minimal length equal to 0, while the right panel is for 
$\epsilon =\frac{l}{m}=0.01$ (see next sub-section). 
The change is not significant, implying that the mere
presence of vacuum fluctuation creates this peak. When $b=0$ (no vaccum fluctuations),
the function $Tm$ is steadily decreasing in $r$, with no peak.

For the entropy, we obtain

\beqa
dS & = &
\left(2 \pi my_h^2\right) dm\left[ 
\left( 1- \frac{a^2 {\rm cos}^2(\theta )}{y_h}
- \frac{b}{3 y_h^2}\right) \left( 1 + \frac{a^2
{\rm cos}^2 (\theta )}{y_h} \right) \right]^{-1}
~~~,
\label{entropy-2a}
\eeqa
which exhibits the same behavior as in the last section. It tends to
infinity toward an event horizon, but because $T$ tends to 0 at the event horizon, the
product $TdS$  stays constant. 

\subsection{The general case: Kerr metric, with $l>0$}
\label{secKerrgeneral}

When $l>0$ another distinct property emerges more pronounced, 
namely negative temperatures. This sounds at 
first strange, because in classical mechanics negative temperatures are not possible.
In classical physics, the temperature is related to the mean kinetic energy of particles 
in a gas. The temperature increases when the  energy of the system increases. However,
this changes in Quantum systems, when there is an upper bound to the energy 
\cite{T-neg,thermo-greiner}. When energy increases, the temperature also increases, until
half of the possible energy, which can be deposited is reached. From that point on,
though the energy increases, the temperature becomes negative. In different words,
negative temperature describes systems which are hotter than systems with a positive 
temperature \cite{T-neg,ramsey}. A central point is that the 
stability relation $\frac{\partial S}{\partial V}$
= $\frac{p}{T}>0$ is valid, where $S$ is the entropy, $V$ the volume, $p$ the pressure and $T$ the
temperature. Thus, when $T$ is negative, the pressure has to be negative, too. Consequently,
at negative temperature there is also a negative pressure, which can stabilize a black hole 
\cite{T-neg}. A useful reference on negative temperature and measurements in real
physical systems is given in \cite{braun,braun-science}.

In this context, we discuss the effects when a minimal length is included.
The infinitesimal length squared changes to \cite{iwara2022,AN2024}

\beqa
d\omega^2 & = & \sigma^2 (r) g_{\mu\nu} dx^\mu dx^\nu
~~~,
\label{l-eq-1}
\eeqa
with

\beqa
\sigma^2 (y) & = & \left( 1 - l^2 \mid a\mid^2\right) 
~~~.
\label{l-eq-2}
\eeqa
where $\mid a \mid^2$ is the absolute value of the acceleration.
As noted by \cite{feoli}, the presence of an acceleration breaks 
covariance. The extraction of the factor $\sigma^2$
has to be seen as an approximation in order to include 
acceleration effects.
Thus, the $\sigma^2(y)$ is a function of the acceleration. In terms of the accelerations
it is given by

\begin{eqnarray}
\sigma ^{2}(r) & = & 1-l^{2}\left[ \left( 1-\frac{\Psi }{y^{2}}\right) \left( 
\overset{..}{t_{0}}\right) ^{2}-\frac{y^{2}}{\Delta }\left( \overset{..}{y}
\right) ^{2}
\right.
\nonumber \\
&& \left.
-\left( \left( y^{2}+\alpha ^{2}\right) +\frac{\alpha ^{2}\Psi }{
y^{2}}\right) \left( \overset{..}{\phi }\right) ^{2}+\frac{2\alpha \Psi }{
y^{2}}\left( \overset{..}{t_{0}}\overset{..}{\phi }\right) \right]
~~~.
\label{l-eq-3}
\end{eqnarray}
with

\begin{eqnarray}
\Sigma &=&y^{2}+a^{2}\cos ^{2}\theta \\
\Delta &=&y^{2}+a^{2}-\Psi \\
\Psi &=&2y-\frac{b}{6y^{2}}
\end{eqnarray}

The accelerations $\ddot{t}$, $\ddot{y}$ and $\ddot{\phi}$ are given in \cite{iwara2022}
and will not be listed here.

Again, the Hawking temperature is given by $T =\frac{\kappa}{2\pi}$, where $\kappa$ is proportional
to the derivative of $g_{00}$, as defined before. 
As we saw, this temperature, times the mass $m$, 
can be defined as
a function in $y$, implying that also at space positions different to the event horizon, there
exists the possibility to produce pairs of particle-antiparticles, just like the mentioned
{\it Schwinger effect} \cite{gelis2016} in Quantum Electrodynamics. 
Thus, when we plot the Hawking temperature as a 
function in $y$, we can access important information on particle pair production at any point 
in space. The dependence of the temperature on the position in the gravitational field
was already investigated in \cite{tolman}. 
The temperature has two terms, one proportional to $\sigma^2$ and the other one
to the derivative of it. I.e., when $\sigma^2< 0$, the temperature becomes negative,
which is only valid if this happens for $y \geq y_h$. This is a more
dominant effect because $\sigma^2$ can have singularities, even for $y \geq y_h$.
On the right hand side of Fig. \ref{figcut} the result for including $\sigma^2$ is shown,
which is similar to \cite{falke2023}, showing the advantage of our phenomenological path, compared
to first principle approaches.

\begin{figure}
\rotatebox{0}{\resizebox{125pt}{125pt}{\includegraphics[width=0.23\textwidth]{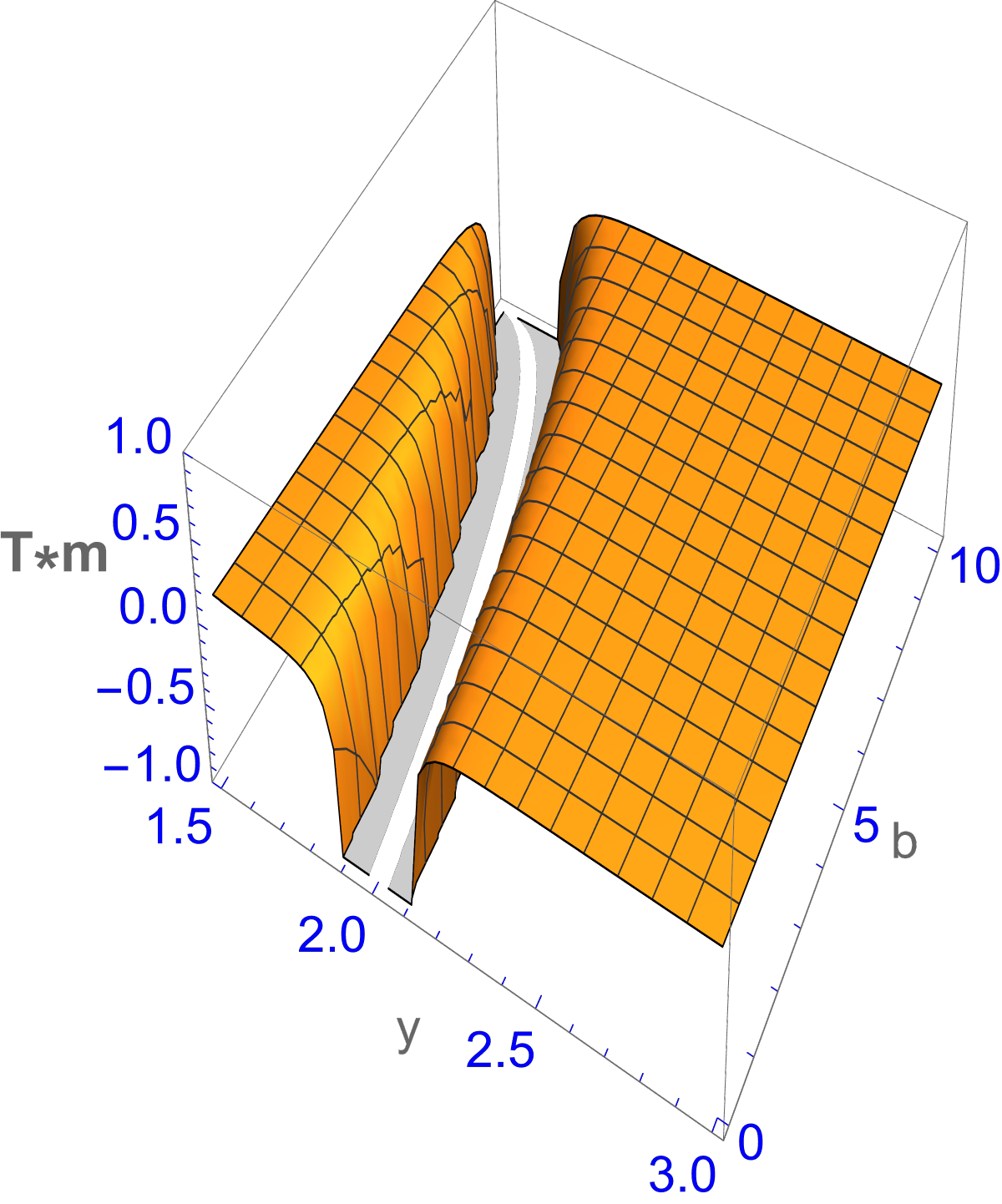}}}
\rotatebox{0}{\resizebox{125pt}{125pt}{\includegraphics[width=0.23\textwidth]{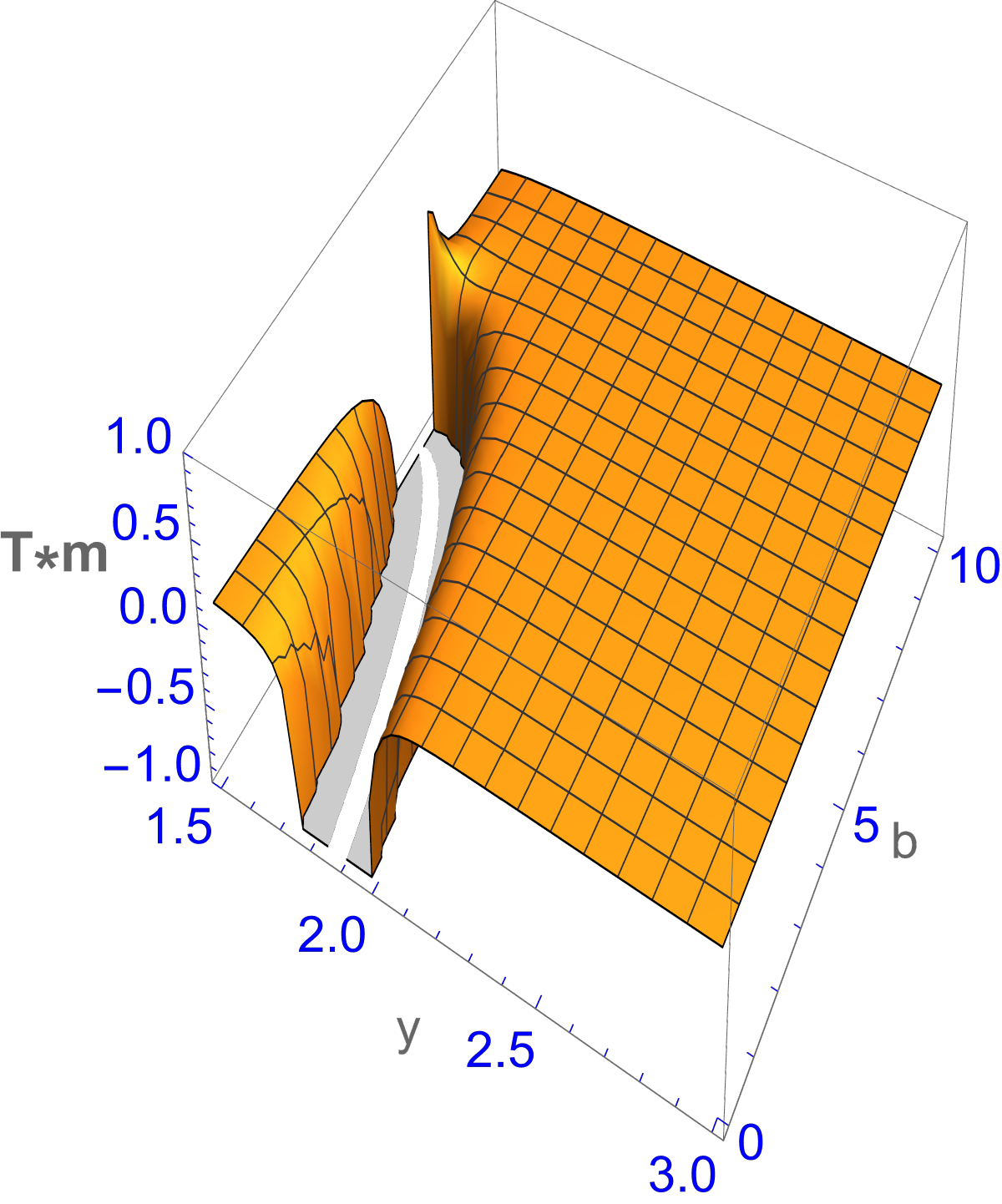}}}
\rotatebox{0}{\resizebox{125pt}{125pt}{\includegraphics[width=0.23\textwidth]{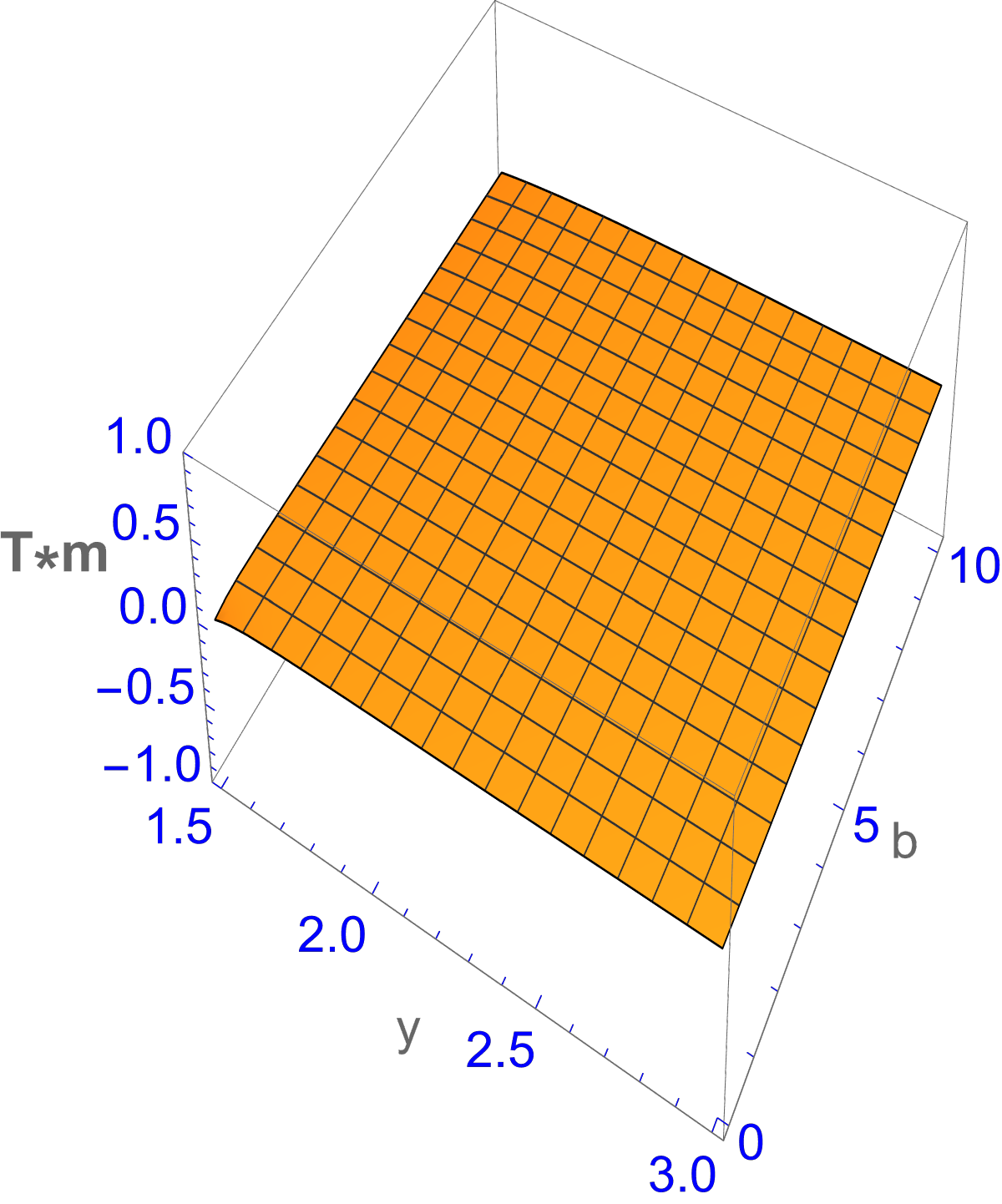}}}
\caption{The Hawking temperature time $m$ as a function on $b$ and $y$. The most left figure
depicts the results for $a=0$, the centered one for $a=0.5$ and the right one for $a=1$. 
We used $a=0$ and 
$\epsilon = \frac{l}{m} = 0.01$, i.e., a black hole mass which
is 100 times larger than the Planck mass. The structure below $y_h$ can be ignored, 
because there is mass present. 
}
\label{Ka}
\end{figure}

In Fig. \ref{Ka} the Hawking temperature is plotted versus $y$ and $b$ for three
different values of $a$ (0, 0.5 and 1). For $a=0$ we note a ridge still appearing at
$y > y_h$ (the structure at $y<y_h$ can be ignored, because we do not 
include mass distributions), representing a singularity,
which follows the position of the event horizon, starting at $y=2$ for $b=0$ and ending
at $y=1.5$ for $b=\frac{81}{8}$. When increasing $a$ the end of the ridge moves to lower values of 
$b$ and vanishing for $a=1$, where no event horizon exists anymore.

Another feature is the appearance of a singularity, where the temperature lowers to
infinite negative values, for $y \geq y_h$. 
As exposed in \cite{T-neg}, negative temperatures correspond
to negative pressure, i.e. it works against the gravitational attraction and stabilizes the
black hole. This was verified experimentally, simulating a black hole by a
Ising-spin systems \cite{exp-BH}. There it is shown that the Ising spin system 
simulates a Schwarzschild black hole, where the strength of the applied
magnetic field is equivalent to the mass \cite{exp-BH}. 
In pcGR this is the consequence of the presence
of dark energy, acting repulsively.

\begin{figure}
\begin{center}
\rotatebox{0}{\resizebox{200pt}{200pt}{\includegraphics[width=0.23\textwidth]{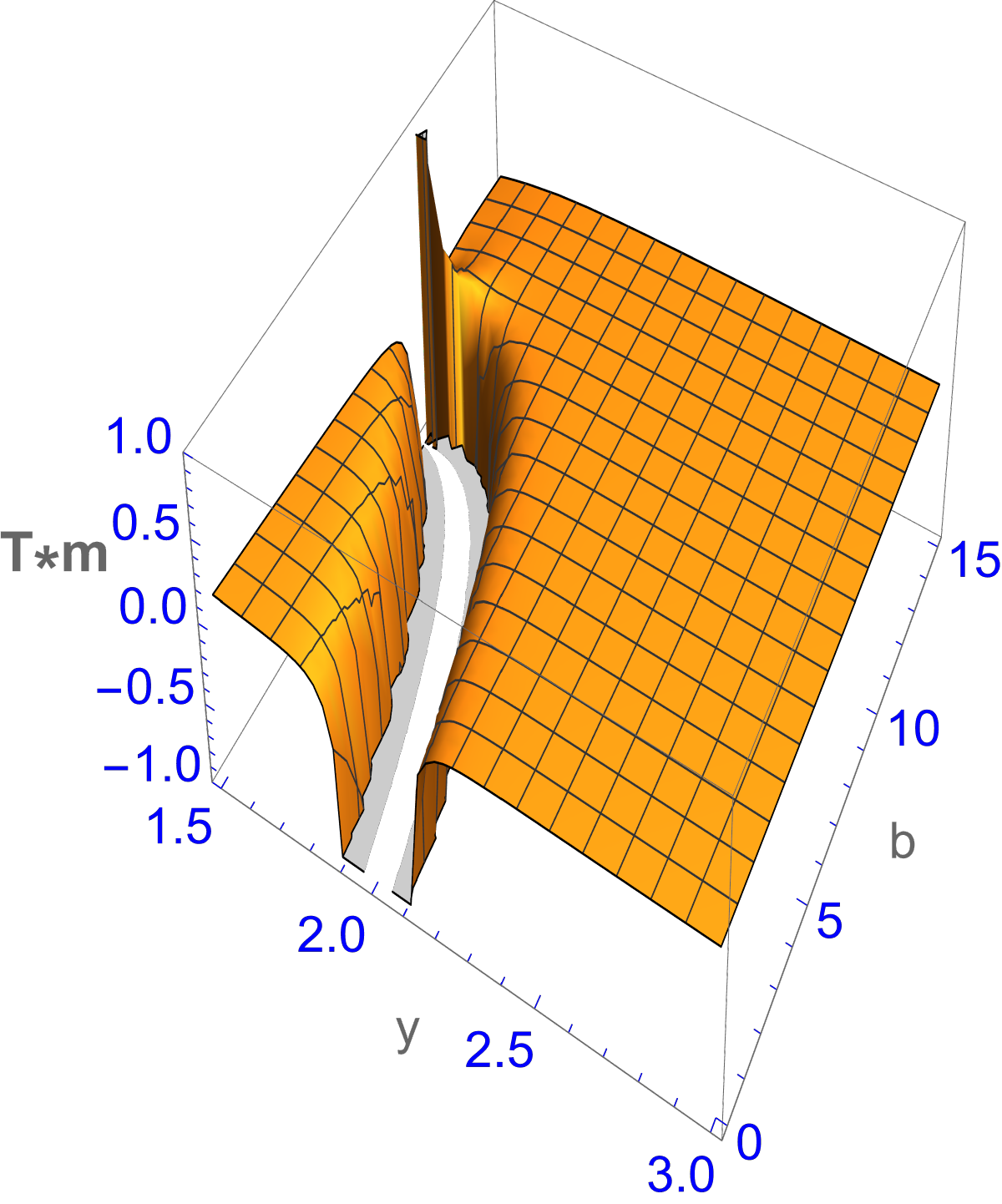}}}
\end{center}
\caption{The Hawking temperature times $m$ as a function on $b$ and $y$, for a=0 but
$b$ ranges up to 15. As can be appreciated, for $B>10.125$, when no event horizon exists, the
temperature does not show a singularity.
We used $a=0$ and 
$\epsilon = \frac{l}{m} = 0.01$, i.e., a black hole mass which
is 100 times larger than the Planck mass. The structure below $y_h$ can be ignored, 
because there is mass present. 
}
\label{Ka-2}
\end{figure}

In Fig. \ref{Ka-2} the Hawking temperature times $m$ is plotted up to $b=15$, which is
above the value of $\frac{81}{8}$, where the last event horizon exists for $a=0$.  As can be appreciated, the temperature does not exhibit a singularity as soon as no event horizon exists,
in agreement to the Fig. \ref{Ka}, where above a specific $a$ value, from which on no event
horizon exists, the temperature also does not exhibit a singularity. 

One of the main take-aways is when an event horizon is present, the Temperature 
rises to very large negative values for $y$ near above the event horizon. 
Negative temperatures are {\it hotter} than 
positive temperatures \cite{T-neg} 
and the result implies an immediate evaporation of
the black hole. 
All this is only valid for small masses, of few orders of the Planck mass, and disappears
for large masses. The implication is that the production of small black holes,
as speculated, is highly suppressed. Thus, according to our theory, black holes can 
only form when they are rapidly produced and with a large mass. 

\section{Conclusions}
\label{con}

In this contribution we resumed the pcGR and discussed their consequences, which are i) the 
existence of a minimal length, ii) the appearance of a dark energy-momentum tensor on the
right hand side of the Einstein equations and iii) some experimental predictions.
The concept of a {\it generalized Mach's principle} is presented, implying that a mass
not only curves space-time but also modifies the vacuum structure around that mass.

We then investigated the consequences of a minimal length for the Hawking
temperature and the entropy in the general case
of a rotating stellar object (Kerr metric) with a minimal length present.
The results presented include GR and theories which inlcude dark energy.

For massive objects, as a massive star or black hole, no effects of the minimal
length are found. The situation changes significantly when the stellar object is small,
like a mini-black hole. Thus, the consequences obtained here are rather valid only for very small
objects, of a few orders of the Planck mass. Therefore, it may only play a role
at the beginning of the universe where such objects are assumed to be produced. 

At the event horizon, the Hawking temperature decreases with increasing accumulation of
dark energy, described by a parameter $b$ in the theory. There is a maximum value ($\frac{81}{8}$)
where the temperature reaches zero (no emission) at the event horizon 
and the entropy tends to infinity,
however, for larger distances the temperature is again positive and shows a maximum 
at a finite radial distance.
This singularity is a consequence of the
simplified assumption of the dark energy distribution.
When the minimal length is present, the metric involves a factor $\sigma^2$, which depends on
the acceleration of a particle. This function shows a singularity at the event horizon, 
when $g_{00}=0$, which This produces negative values in the Hawking temperature, tracing
the position of the event horizon. As soon as $b>\frac{81}{8}$, no large negative temperatures appears
and the Hawking temperature has finite values. Small black holes evaporate explosively due to
the appearance of a negative temperature at an event horizon. Thus, their
very production is strongly suppressed. It appears that negative temperatures only 
appear  
due to the event horizon, because when no event horizon exists, the temperature remains finite.
Negative temperatures also imply a negative pressure \cite{T-neg}, which is explained in pcGR 
by the presence of dark energy, stabilizing the black hole. There is also the case when $T=0$, 
which implies that there is no radiation and the star is stable. 

These findings are only valid for very small stellar
objects. According to our findings, black holes have to be produced rapidly and with large
masses, in order to survive.

\section*{Acknowledgements}

L.M. and P.O.H. acknowledge financial support by DGAPA-PAPIIT (IN116824).


\begin{thebibliography}{99}

\bibitem{pcGR-first} P. O. Hess, W. Greiner, Int. J. Mod. Phys. E 
{\bf 18} (2009), 51.

\bibitem{book} P. O. Hess, M. Sch\"afer, W. Greiner, {\it Peudo-Complex General Relativity},
(Springer, Heidelberg, 2015).

\bibitem{PPNP} P. O. Hess, Alternatives to Einstein's Relativity Theory, 
Prog. in Part. and Nucl. Phys. {\bf 114} (2020) 103809.

\bibitem{accr} P. O. Hess, AN {\bf 342} (2021),735.

\bibitem{hess-light-poles} Hess, publ. where light emission near the poles is predicted.


\bibitem{schwinger-effect} Y. Kluger, J. M. Eisenberg, B. Svetitsky, F. Cooper, E. Mottola,
Physical Review Letters {\bf 67} (1991), 2427.

\bibitem{T-neg} R. A. Norte, EPJ {\bf145} (2024), 29001.

\bibitem{kelly} P. F. Kelly, R. B. Mann, 
Ghost properties of algebraically extended theories of gravitation,
Class. and Quant. Grav. {\bf 3}, 705 (1986).

\bibitem{birrell} N. D. Birrell, P. C. W. Davis, {\it Quantum Fields
is Curves Space}, Cambridge University Press, London, (1982).

\bibitem{visser} M. Visser, Phys. Rev. D 54 (1996) 5116.

\bibitem{adler} R. Adler, M. Bazin, M. Schiffer, {\it Introduction to General Ralativity}, 2nd edition,
(Mcgrw-Hill, New York, 1975).

\bibitem{caianiello} E. R. Caianiello, Il Nuovo Cim. Lett. {\bf 32} (1981), 65.

\bibitem{feoli} A. Feoli, G. Lambiase, G. Papini, G. Scarpetta 2000, Phys. Lett. A {\bf 268}
(2000), 247.

\bibitem{caspar} G. Caspar, I. Rodr\ii guez, P. O. Hess, W. Greiner,
Int. J. Mod. Phys. E {\bf 25} (2016), 1650027.

\bibitem{padman} T. Padmanabhan, Phys. Rep. {\bf 406} (2005), 49.

\bibitem{falke2023} M. F. Wondrak, W. D. van Suijlekom, Heino Falcke, Phys. Rev. Lett.
{\bf 130} (2023), 221502.

\bibitem{symmetries-2020} P. O. Hess, E. López-Moreno, 
Universe {\bf 5} (2019), 191.

\bibitem{gelis2016} F Gelis, N Tanoi, Progr. Part. Nucl. Phys. {\bf 87} (2016), 1.

\bibitem{thermo-greiner} W. Greiner, L. Neise, H. Stoecker, {\it Themrmodynamics and
Statistical Mechanics}, (Springer Heidelberg, 1995).

\bibitem{ramsey} N. F. Ramsey, Phys. Rev. {\bf 103} (1956), 20. 

\bibitem{braun} S. Braun, {\it Negative Absolute Temperature and the Dynamics of
Quantum Phase Transitions}, PhD thesis, Fakult\"at f\"ur Physik
der Ludwig-Maximilians-Universit\"at M\"unchen, Germany (2014).

\bibitem{braun-science} S. Braun,1, J. P. Ronzheimer, M. Schreiber, S. S. Hodgman, 
T. Rom,1, I. Bloch, U. Schneider, Science {\bf 339} (January 4) (2013), 52.

\bibitem{iwara2022} L. Maglahoui, P. O. Hess, AN {\bf 344} (2022), 220067.

\bibitem{AN2024} L. Maghlaoui, P. O. Hess, AN {\bf 345} (2024), 20230151. 

\bibitem{tolman} R. C. Tolman and P. Ehrenfest, Phys. Rev.  {\bf 36} (1930), 1791.

\bibitem{exp-BH} Oppenheim J., Phys. Rev. E {\bf 68} (2003), 016108.

\end{thebibliography}
\end{document}